\documentclass{ws-procs9x6}
\begin{document}

\title{Unidentified EGRET sources in the Galaxy}

\author{Isabelle A. Grenier}

\address{Universit\'{e} Paris VII \& CEA Saclay\\
Service d'Astrophysique, 91191 Gif/Yvette, France\\
E-mail: isabelle.grenier@cea.fr}

\maketitle

\abstracts{Identifying $\gamma$-ray sources in the Galaxy is
hampered by their poor localization, source confusion, and the
large variety of potential emitters. Neutron stars and their
environment offer various ways to power $\gamma$-ray sources:
pulsed emission from the open magnetosphere and unpulsed $\gamma$
rays from the wind nebula and from the cosmic rays accelerated in
the supernova remnant. While the latter still awaits confirmation,
new candidate associations bring forward the importance of
$10^{4}$ yr old pulsars as GeV sources, with a diversity that will
help constrain the acceleration mechanisms near the pulsar and in
the wind. Theoretical interest in the $\gamma$-ray activity of
X-ray binaries and micro-quasars has also been revived by the
emergence of a subset of variable sources in the inner Galaxy and
another one in the halo.}

Only $\sim$80 of the 271 EGRET sources\cite{hartman99} have been
identified because of the large position uncertainty resulting
from the poor angular resolution and the structured interstellar
background. Irrefutable identifications with a flaring blazar or a
pulsar are even fewer. Most identifications are blazars, based on
spatial coincidence with a bright, flat-spectrum radiogalaxy.
Their number agree with the total expected from the largely
anisotropic distribution of all sources across the sky. So, nearly
two thirds of the sources are hiding in our Galaxy, awaiting
identification.

\section{In the Galactic disc}\label{lowlat} A population
of rather hard and steady sources is found at $|b|<3^\circ$ along
the Galactic plane, mostly in the inner Galaxy. Their average
spectral index is $\bar{\gamma}= 2.09 \pm 0.02$ and 80\% of them
show no sign of time variability\cite{tompkins99}. The positional
correlation with tracers of star formation, such as HII regions,
pulsars, supernova remnants (SNR), and OB associations, points to
an origin in or near active star-forming sites. The latest
compilation\cite{romero99} yields combined probabilities for
chance alignment with 23 OB associations and 26 SNRs below
$10^{-3}$ and $10^{-5}$, respectively. Distances of 1 to 2.5 kpc
for the OB associations\cite{romero99} and of 1 to 4 kpc from the
source spatial distribution imply luminosities $L_{>100 MeV,4\pi
sr}$ of (0.6-4) $10^{28}$ W.

The stable pulsars show a standard deviation in flux of 10-12\%
that is attributed to systematic uncertainties in the instrument
performance. Unidentified sources are found to vary less at lower
latitude\cite{nolan03}. Yet, 17 sources along the Galactic plane
exhibit a lower limit in flux standard deviation that is twice as
large as the instrumental dispersion\cite{nolan03}. These sources
vary over months and are concentrated at $|l| \le 55^\circ$, i.e.
in the inner spiral arms. Distances of 5 to 8 kpc imply high
luminosities $L_{>100 MeV,4\pi sr}$ of (0.5-5) $10^{29}$ W that
are typical of X-ray binaries and $10^{4}$-yr-old pulsars. The
latter are expected to be steady emitters, but variability over
months is common in accreting systems. The large dispersion in
spectral indices, ranging from 1.7 to 3.1, yields no further clue.

\section{In the Gould Belt}\label{subsec:gould} The stable
sources gathering at $3^\circ <|b|< 30^\circ$ significantly differ
from those in the inner Galaxy. They exhibit softer spectra and
lower fluxes with a distinctly steeper $log(N)-log(>S)$
distribution than near the plane\cite{gehrels00}. Their excess at
mid latitudes points to a nearby origin and closely follows the
trace of the Gould Belt around us. It is significantly better
correlated with the Belt than with other Galactic structures
\cite{grenier00}. As a 30 to 40 Myr-old starburst region, 300 pc
in radius, the Belt has recently produced nearby supernovae at a
rate 3 to 5 times higher than the local Galactic one
\cite{grenier00}. As many as $45 \pm 6$ persistent sources can be
associated with the Belt, among which $\sim10$ may be background
sources in the Galactic disc. They have luminosities $L_{>100
MeV,4\pi sr}$ of (0.3-8) $10^{26}(D/300 pc)^{2}$ W. Most error
boxes lack suitable counterparts, but neutron star activity offers
a promising prospect, in particular emission from Myr-old pulsars
\cite{grenier00,hardingzhang01}.

\section{At large scale height} Variable sources at $|b|>3^\circ$
form a separate population with spectral, temporal, and spatial
properties different from the persistent
sources\cite{grenier00tonan}. It is unlikely that the difference
be due to systematic biases in the instrument performance, the
survey exposure, or the Galactic background. Average spectral
indices of $2.25 \pm 0.03$ and $2.52 \pm 0.06$ were obtained for
the persistent (p) and non-persistent ($\bar{p}$) sources, with a
chance probability of 2 $10^{-7}$ of equal
index\cite{grenier00tonan}. Average variability
indices\cite{tompkins99} of $0.38 \pm 0.06$ and $0.95 \pm 0.18$
were obtained with a chance probability of 1.3 $10^{-4}$ of equal
$\tau$. The $\bar{p}$ sources behave much like the variable EGRET
blazars and the p sources show no or little
variability\cite{nolan03}. While the p sources gather along the
Gould Belt, the $\bar{p}$ sources are scattered within $60^\circ$
around the Galactic center, with a 4 $10^{-8}$ probability of
identical distribution. The $45 \pm 9$ $\bar{p}$ sources are not
isotropically distributed ($4.7 \sigma$), but have a large
scale-height $z_{H} = 2.0^{+1.2}_{-0.6}$ kpc above the Galactic
plane or a radial distribution in the halo equivalent to that of
the globular clusters (Grenier, in prep.).  Their location implies
luminosities $L_{>100 MeV,4\pi sr}$ of $10^{28-30}$ W and ages of
the order of $10^{9}$ yr.

\section{Massive stars}
Particles accelerated in the supersonic wind of massive stars, at
the terminal shock or along the turbulent $10^{28-29}$ W wind, can
up-scatter the stellar UV radiation field to produce $\gamma$
rays. Yet, none of the numerous nearby O stars in the Gould Belt
has been detected by EGRET\cite{grenier00}. Binary systems of
massive stars should be more efficient by providing ample UV
target photons and a strong shock between colliding winds.
Synchrotron radio emission has indeed been observed in several
systems. The WR+O star system W140 has been proposed to explain
the stable 3EG J2022+4317 source\cite{benaglia02} and the O+O+B
system Cyg OB2 n$^\circ$5 to account for half of the 3EG
J2033+4118 flux\cite{benaglia01}. Collective acceleration from
stellar winds and nearby supernova shocks, as in
SNOBs\cite{montmerle79} or in
superbubbles\cite{bykov01superbubbles}, could also produce
extended $\gamma$-ray sources that EGRET could not resolve.

\section{X-ray binaries and micro-quasars}
In high-mass X-ray binaries, electrons accelerated at the shock
between the pulsar and the stellar winds can shine up to TeV
energies\cite{tavaniarons97,kirk99} by up-scattering the stellar
radiation. PSR B1259-63 has been extensively studied near
periastron when TeV synchrotron emitting electrons are produced.
But, the emission peak should occur at TeV energies, explaining
the lack of EGRET detection\cite{kirk99}. The association of SAX
J0635+0533 with the hard and stable 3EG 0634+0521
source\cite{kaaret00} would offer a unique opportunity to study
the early phase of a massive system hosting a young (1.4 kyr) and
energetic (5 $10^{31}$ W) pulsar. Alternatively, if a transient
accretion forms and rotates more rapidly than the pulsar, protons
may be accelerated to TeV energies in the magnetosphere, near the
null surface. Colliding further out with the disc, they would
produce $\pi^{0}$-decay $\gamma$ rays. This scenario could explain
the soft, $E^{-2.67 \pm 0.22}$, variable 3EG 0542+2610 source
coincident with the X-ray transient A0535+26\cite{romero01}.

In micro-quasars, energetic electrons in persistent jets can
up-scatter accretion disc photons or stellar radiation to a few
MeV for a low-mass companion and to GeV energies for a high-mass
one, depending on the jet viewing angle\cite{georganopoulos02}.
Synchrotron-self-Compton emission is fainter. Sporadic ejections
would shine too briefly to account for EGRET sources. Tidal
precession would cause variability on a few month timescale, so
micro-quasars with high-mass companions have been proposed to
explain variable sources in the inner Galaxy\cite{kaufman02}. The
LS 5039 system\cite{paredes00} is indeed seen toward 3EG
J1824-1514, but the source is stable. Whether the low-mass systems
can explain the variable halo sources is being explored.

3EG 0241+6103 has long been associated\cite{kniffen97} with the
radio source LSI $+61^\circ$ 303 which is known for its radio
flares after periastron, its mildly relativistic jet and the 4-yr
precession of its disc\cite{massi01}. But the EGRET variability
does not correlate with the radio phase and the COMPTEL flux is
stable. So, the association is unclear.

\section{Pulsars and their wind nebulae}
Hard and stable sources are often related to pulsars following the
characteristics of the 7 young and energetic EGRET pulsars. But
older, fainter pulsars may behave differently. Pulsar wind nebulae
(PWNe), confined by the interstellar pressure or by the ram
pressure from the pulsar motion, have also emerged as potential
candidates up to TeV energies as in the Crab, Vela, and PSR
J1709-4429 nebulae. 7 EGRET sources coincide with PWNe. In
particular, PSR J1048-5832 shows unpulsed GeV emission possibly
arising from the wind nebula\cite{kaspi00}.

Six of the 9 radio pulsars with highest $\dot{E}/D^{2}$ rank are
seen in $\gamma$ rays, indicating a close relationship between the
onset of high-energy showers and coherent radio emission.
Population studies using the characteristics of the $\gamma$-ray
beam from the polar cap and the outer gap models, as well as the
radio beam properties, show that pulsars likely dominate the
unidentified source population at low
latitude\cite{gonthier02,zhangzhangcheng00}. Polar-cap and
outer-gap pulsars can account for $\sim$60\% and $\sim$100\% of
the stable sources, respectively. The two models predict different
source counts and distinct ratios of radio-loud to radio-quiet
$\gamma$-ray pulsars because of different beam apertures, a
tighter correlation between the radio and $\gamma$-ray beams above
the polar cap, and very different evolutions with age. The polar
cap (or outer gap) simulations yield 13 (or 10) radio-loud and 2
(or 22) radio-quiet pulsars above the EGRET
sensitivity\cite{gonthier02,zhangzhangcheng00}. These differences
and the detection of older pulsars will serve as important
diagnostics for modelling the pulsed emission.

Pulsars also largely contribute to the Gould Belt
sources\cite{grenier00,hardingzhang01} because of the enhanced
supernova rate. Vela and Geminga are the first two examples of
Belt pulsars. Off-beam emission is expected in polar cap models,
due to high-altitude curvature radiation from primary electrons.
This widely beamed emission is softer and fainter than on-beam
emission, much like the Belt sources. Evolving neutron stars born
in the expanding Belt over the last 5 Myr, using both on and off
beams with free apertures and luminosities reflecting the
$L_{\gamma} \propto \dot{E}^{1/2}$ relation observed for the known
pulsars, shows that enough pulsars remain visible to retain the
Belt spatial signature and to account for half of the Belt
sources. They are quite older, thus fainter, than the pulsars
detectable in the Galactic plane (1.5 Myr vs. 0.35 Myr) and many
would be radio quiet (Perrot et al., in preparation).

\begin{table}[ph]
\tbl{radio pulsars coincident with EGRET sources. v notes variable
sources} {\footnotesize
\begin{tabular}{|l|l|l|l|l|l|l|l|} \hline source &sp. index
&pwn &{pulsar} &$D_{NE2001}$ &age
&{$\dot{E}$} &{$L_{1sr}/\dot{E}$} \\
3EGJ & & &J &kpc &kyr &$10^{28}$W  & \\
\hline 1013-5915 &$2.32 \pm 0.13$ &pwn &1016-5857 &7.6
&21 &25.9 &0.08 \\
1014-5705 &$2.23 \pm 0.2$ & &1015-5719 &5.0
&39 &8.27 &0.05 \\
1102-6103 &$2.47 \pm 0.21$ & &1105-6107 &4.9
&63 &24.8 &0.06 \\
1410-6147 &$2.12 \pm 0.14$ &pwn &1412-6145 &7.7
&51 &1.24 &4.0 \\
1410-6147 &$2.12 \pm 0.14$ & &1413-6141 &9.9
&14 &5.65 &1.0 \\
1420-6038$^{v}$ &$2.02 \pm 0.14$ &pwn &1420-6048 &5.6
&13 &104 &0.03 \\
1639-4702 &$2.5 \pm 0.18$ & &1637-4642 &5.0
&41 &6.40 &0.15 \\
1714-3857 &$2.3 \pm 0.2$ & &1715-3903 &4.1
&117 &0.689 &1.0 \\
1824-1514 &$2.19 \pm 0.18$ & &1825-1446 &5.0
&195 &0.412 &1.7 \\
1837-0423$^{v}$ &$2.71 \pm 0.44$ &pwn &1838-0453 &7.9
&52 &0.827 &9.9 \\
1837-0606 &$1.82 \pm 0.14$ & &1837-0604 &6.3
&34 &20.0 &0.08 \\
1856+0114$^{v}$ &$1.93 \pm 0.1$ &pwn &1856+0113 &3.1
&20 &4.30 &0.08 \\
2021+3716 &$1.86 \pm 0.1$ &pwn &2021+3651 &12.1
&17 &33.8 &0.18 \\
2227+6122 &$2.24 \pm 0.14$ &pwn &2229+6114 &7.2
&10 &225 &0.04 \\
\hline
\end{tabular} }
\vspace*{-13pt}
\end{table}

The sample of radio pulsars has more than doubled since the end of
EGRET, but it was not possible to search back for pulsations in
the $\gamma$-ray data. Out of 1412 pulsars in the ATNF catalogue,
36 coincide with unidentified sources, but only the 14 listed in
Table 1 exhibit $\gamma$-ray luminosities over 1 sr below or close
to the pulsar spin-down power $\dot{E}$, allowing for some
distance uncertainty. The candidate counterparts span similar ages
of $10^{4-5}$ yr and $\dot{E}$ of $10^{28-30}$ W than the known
$\gamma$-ray pulsars, but they are much more distant and notably
softer. The larger distances result in large
$L_{\gamma,1sr}/\dot{E}$ efficiencies well in excess of the
observed $L_{\gamma,1sr} \propto \dot{E}^{1/2}$ relation, unless
their beams are much narrower or the pulsed fraction is low.

PSR J2229+6114, second only to the Crab in $\dot{E}$, is a
compelling identification for the stable EGRET
source\cite{halpern01}. Were it not, one should explain an
unprecedented flux ratio $f_{\gamma}/f_{X} > 10^{4}$. Coincident
with a soft COMPTEL source, its spectral distribution would peak
in the MeV band as in Vela. It also produces an X-ray jet and an
equatorial wind which powers a compact, bow-shock like PWN. PSR
J1016-5857 is surrounded by an X-ray wind nebula and its radio
wake may extend to the SNR G284.3-1.8 at 3 kpc\cite{camilo01}. Its
efficiency at this distance would compare with that of the
$\gamma$-ray pulsars PSR J1709-4429 and J1048-5832 of equivalent
age and $\dot{E}$. PSR J1420-6048, in the Kookaburra nebula, may
be as close as 2 kpc. Its efficiency would be reasonable, but the
source variability suggests the PWN may be active in $\gamma$
rays. The nearby Rabbit plerion could also contribute to the flux.
A pulsar origin of 3EG 1837-0423 is unlikely because of the pulsar
weakness and the source variability. The plerion in G27.8+0.6
offers an alternative. Error boxes are crowded: there is a bright
ASCA source near PSR J1837-0604, the bright WR 141 star near PSR
J2021+3651\cite{roberts02}, and 3 SNRs toward 3EG 1639-4702.
Coincidences with other famous SNRs are discussed below. Both PSR
J1412-6145 and J1413-6141 can hardly produce the $\gamma$-ray
flux, even at a distance a few kpc. PSR J1412-6145 is the likely
progenitor of G312.4-0.4, but too weak to explain the wind-like
nebula\cite{doherty02}.

Radio-quiet pulsars are also promising candidates, but blind
searches for pulsation have failed\cite{chandler01blind}. The
bright 3EG 1835+5918 source displays striking "Geminga-like"
properties\cite{mirabal01,reimer01}, i.e. a stable $E^{-1.73 \pm
0.07}$ spectrum cutting off at 4 GeV, a faint soft X-ray
counterpart with no optical or radio emission, and a flux ratio
$f_{X}/f_{V} > 300$ typical of neutron stars. Its luminosity
$L_{\gamma,1 sr} = 4.6$ $10^{26} (D/1 kpc)^{2}$ W would be 10
times larger than Geminga. 3EG J0010+7309 also exhibits a stable
$E^{-1.58 \pm 0.18}$ spectrum cutting off at 2 GeV and an X-ray
counterpart with no optical emission\cite{brazier98}. The 1.7
$10^{29}$ W compact keV nebula observed inside the CTA1 remnant
could indeed be powered by 20 kyr radio-quiet pulsar. A third case
of a radio-quiet pulsar was proposed in $\gamma$
Cygni\cite{brazier96}.

Another candidate PWN has been found\cite{braje02} toward the
variable, $E^{-2.06 \pm 0.08}$, source 3EG 1809-2328. GeV
electrons irradiating the nearby cloud would not explain the
variability timescale of months. Up-scattering of the ambient
radiation field from the nearby OB association or the nebular
synchrotron emission largely fail to explain the source. The
$L_{>100 MeV,1 sr} = 8.8$ $10^{26}$ W luminosity is typical of a
$10^{4-5}$ yr-old pulsar, but not the variability. So, this source
is extremely puzzling.

Old millisecond pulsars have become attractive counterparts since
the discovery of pulsed $\gamma$ rays $\le 300$ MeV from PSR
J0218+4232. The identification\cite{kuiper02} is based on the
alignment of the X-ray, $\gamma$-ray, and radio pulsed profiles.
The soft emission peaks in the MeV range and is stable. The
distance of 5.7 kpc in the halo implies a luminosity $L_{>100
MeV,1sr} = 1.6$ $10^{27}$ W and a 7\% efficiency that both the
polar-cap and outer-gap models can explain because the weak
surface field ($B = 8.5$ $10^{4}$ T) is compensated by a very
compact magnetosphere. Yet, both models fail to reproduce the
spectral distribution. Whether ms pulsars can explain the equally
soft, but variable halo sources remains an open question.

\section{Supernova remnants}
Supernova shocks accelerate particles to very high energies in a
highly non-linear way. When the particle energy density becomes
significant, the gas is more compressible, the acceleration rate
and the post-shock density increase, and the temperature drops.
So, both thermal X rays and $\gamma$ rays shed light on cosmic-ray
acceleration. TeV electrons are clearly revealed by synchrotron X
rays, but interpreting the $\gamma$ rays in terms of electron and
proton emission critically depends on the highly uncertain
magnetic field at the shock and in the remnant. Electron emission
dominates for a compressed B $\sim$ 1 nT, as in
G347.3-0.5\cite{ellison01RX}. Proton emission may prevail for
non-linear amplification of B up to 10-100 nT by the cosmic rays,
as in Cas A\cite{berezhko03CasA} and SN
1006\cite{berezhko02SN1006}.

Coincidences between EGRET sources and famous SNRs, such as IC443,
W28, W44, $\gamma$ Cygni, CTA1, and G347.3-0.5, have been
reported, but the origin of the emission is not clear. In IC443,
the 3EG 0617+2238 error box clearly exclude the X-ray pulsar and
its plerion. It points to the shell center where no peculiar
activity may explain the stable $E^{-2.01 \pm 0.06}$ source which
does not seem to pulse\cite{chandler01blind}. A $10^{4} M\odot$
cloud may serve as target to the freshly accelerated cosmic
rays\cite{butt02IC443}. Cloud irradiation is also proposed for 3EG
J1714-3857, which lies towards a 3 $10^{5} M\odot$ cloud next to
the G347.3-0.5 shell\cite{butt02RX}, and for sources found on the
rim of old nearby radio shells at medium
latitudes\cite{combi98,combi01}. In W44, electrons pervading a
cloud or up-scattering the SNR radiation can reproduce the
$\gamma$-ray flux\cite{dejager}. But the source coincides with PSR
J1856+0113 and its bow-shock PWN that shines in synchrotron up to
keV energies. The source spectrum is typical of a pulsar, but not
its variability. So, the PWN offers an attractive alternative.

In short conclusion, $\gamma$-ray data with finer angular
precision and ample statistics to search for variability and
pulsation are desperately needed!

\end{document}